\documentclass{article}
\usepackage[T1]{fontenc}
\usepackage[utf8]{inputenc}
\usepackage{authblk}
\usepackage{lmodern}
\usepackage{amsmath}
\usepackage{amsfonts}
\usepackage{braket}
\usepackage{amssymb}
\usepackage{braket}
\usepackage[caption = false]{subfig}
\usepackage{graphicx}
\usepackage[left=2.54cm, right=2.54cm, top=2.54cm, bottom=2.54cm]{geometry}
\usepackage{float}
\usepackage{notoccite}
\usepackage{hyperref}
\usepackage{titlesec}
\usepackage{setspace}
\usepackage{acronym}

\usepackage{indentfirst}

\usepackage{verbatim}

\usepackage{booktabs}

\title{Investigation of Quark Distributions in a Family of Pentaquarks using the Thomas-Fermi Quark Model}

\author[a,b,c]{Mohan Giri\thanks{mohan\_giri1@baylor.edu}}
\author[b,d,e]{Suman Baral\thanks{suman@neuralinnovations.io}} 
\author[b,c]{Gopi Chandra Kaphle\thanks{gck223@gmail.com}}
\author[b]{Nirmal Dangi\thanks{nirmal.dangi2@gmail.com}}
\author[a,b]{Sudip Shiwakoti\thanks{sudip\_shiwakoti1@baylor.edu}}
\author[d]{Leonardo Bardomero\thanks{josebardomero@gmail.com}}
\author[a]{Paul Lashomb\thanks{paul\_lashomb@baylor.edu}}
\author[a,b]{Walter Wilcox\thanks{walter\_wilcox@baylor.edu}}
\affil[a]{Department of Physics, Baylor University, Waco, TX USA 76798}
\affil[b]{Everest Institute of Science and Technology, Kathmandu, NP 44618}
\affil[c]{Central Department of Physics, Tribhuvan University, Kirtipur, NP 44600}
\affil[d]{Neural Innovations LLC, Lorena, TX USA 76655}
\affil[e]{Division of Health Science and Mathematics, Niagara County Community College, Sanborn, NY USA 14132}

\begin{document}
\bibliographystyle{unsrt}
\date{}
\maketitle
\begin{abstract}
Using the Thomas-Fermi quark model, a collective, spherically symmetric density of states is created to represent a gas of interacting fermions with various degeneracies at zero temperature. Over a family of pentaquarks, $uudc\bar{c}$, color interaction probabilities were obtained after averaging over all the possible configurations. Three different functions are developed for light, charm, and anti-charm quarks and are assumed to be linearly related by some proportionality constants. Interesting patterns of quark distributions are observed while analyzing the quark function consistency conditions for such constants.
\end{abstract}

\section{Introduction}
Over the last two decades the existence of multi-quark states such as pentaquarks and tetraquarks have been confirmed through the efforts of the LHCb\cite{Aaij:2015tga,Aaij:2016phn,Aaij:2016ymb,Aaij:2019vzc}, BESIII\cite{Ablikim:2013mio,Ablikim:2015tbp}, Belle\cite{Choi:2003ue,Choi:2007wga,Belle:2011aa,Liu:2013dau} and other collaborations. One can expect even more to be discovered in the years to come. The measurements of the state productions by the LHCb  \cite{Aaij:2019vzc} were determined to be pentaquark states of quark flavor content $uudc\bar{c}$. Specifically, the measured charmonium-pentaquarks were the $P_C(4312)^+$, $P_C(4440)^+$, and $P_C(4457)^+$.

One of the standard theoretical methods to investigate such multi-quark states is Lattice Quantum Chromodynamics (LQCD). As the quark content increases, however, LQCD becomes more computationally expensive and time-intensive. A great deal of effort on theoretical setup in terms of Wick contractions, operator selection, wave function smearing and analysis must be done. Each state must be investigated separately and no global picture emerges. Larger systems also require larger lattices. In order to investigate the dynamics of such exotic states, the Thomas-Fermi (TF) statistical quark model was developed \cite{Wilcox:2008vd} as an inexpensive alternative. We have pointed out that it could be key to identifying {\it families} of bound states, rather than individual cases\cite{Wilcox:2018tau}. The TF quark model has previously been applied to systems of multi-quarks to investigate ground state properties of baryons\cite{Liu:2012my,liu2019thomas}. It has also been used to examine the bound states of multi-quark mesons\cite{Baral:2019qpl}. The latter paper suggested the existence of several tetra, octa, and hexadeca quark states. Due to its timely experimental significance, in this paper we begin a theoretical investigation of the quark distributions for the family of multi-charmonium pentaquarks. These would include penta (5), deca (10), pentedeca (15),... collections of quarks and antiquarks, where we require one additional $c\bar{c}$ combination per penta addition.

The paper is organized as follows. We first develop a formalism to count the number of color interactions in Section \ref{pote}. In Section \ref{flav} we put flavor and color interactions together, convert a discrete system of particles into a continuous system, and obtain normalization conditions. After that, we introduce the TF statistical quark model and obtain expressions for potential and kinetic energies in Sections \ref{te}. The energy is minimized by varying the density of states and TF differential equations are obtained in Section \ref{tfe}. The consistency conditions are then formulated in Section \ref{ccp} for various quark contents and degeneracies. We examine the resulting radial distributions of three different types of quarks (light, charmed and anti-charmed) in Section \ref{rd}. Finally, we give conclusions and outline of further work in Section \ref{conclusions}.

\section{Residual Color Coulombic Interactions}\label{pote}

The types of possible color interactions will be six in number, the same as in Ref.~\cite{Baral:2019qpl}; namely, color-color repulsion (CCR), color-color attraction (CCA), color-anti-color repulsion (CAR), color-anti-color attraction (CAA), anti-color-anti-color repulsion (AAR), and finally anti-color-anti-color attraction (AAA). The statistics, however, will differ from before. In this section, we will determine the average number of times a given interaction will occur, perform a cross check on the calculations, and finally see that the system is bound through residual color coupling alone even in the absence of volume pressure. 
  
\subsection{Occurrence of color interactions}
 Due to color confinement, objects made from quarks must be $SU(3)$ color singlet states in order to exist as free particles. Color singlets can be achieved in five different ways, as shown in Table \ref{tab:tab1}. 
        
\begin{table}[ht]
\centering
\begin{tabular}[t]{ccc}
\toprule
Number & Combinations for a color singlet & Representation\\
\midrule
1 & red + blue + green & $rgb$\\
2 & anti-red + anti-blue + anti-green & $\bar{r}\bar{b}\bar{g}$\\
3 & red + anti-red & $r\bar{r}$\\
4 & blue + anti-blue & $b\bar{b}$\\
5 & green + anti-green & $g\bar{g}$\\
\bottomrule\\
\end{tabular}
\caption{Five different ways to obtain a color singlet.}
\label{tab:tab1}
\end{table}%

A pentaquark is a system of either four quarks and an anti-quark or four anti-quarks and a quark. In order for it to be a color singlet, other arrangements are not possible. This requirement for pentaquarks to be color singlets, and the various ways they can form color singlets affects the number of occurrences of color interactions. For example, a system of ten particles could equally be two red, two blue, four green and two anti-green; or two red, three blue, three green, one anti-blue and one anti-green. So, when we talk about the occurrence of color interactions, there is some probability for each occurrence. We calculate such probabilities for a system of $5\eta$ particles in this subsection. For convenience, we refer to a color singlet consisting of five quarks as a $\textit{pocket}$. This means a system  of $5\eta$ particles have $\eta$ pockets in total. Any pocket could equally be one of the six possibilities as shown in the Table~\ref{tab:tab2} 
below.

\begin{table}[ht]
\centering
\begin{tabular}[t]{ccc}
\toprule
Number & Color singlet\\
\midrule
1 & $rbg + r\bar{r}$\\
2 & $rbg + b\bar{b}$\\
3 & $rbg + g\bar{g}$\\
4 & $\bar{r}\bar{b}\bar{g} + r\bar{r}$\\
5 & $\bar{r}\bar{b}\bar{g} + b\bar{b}$\\
6 & $\bar{r}\bar{b}\bar{g} + g\bar{g}$\\
\bottomrule\\
\end{tabular}
\caption{Different ways a color singlet pocket could be formed.}
\label{tab:tab2}
\end{table}%

In this paper, we are investigating structures of a family of hidden charm multi-pentaquarks. So, our system of fermions can have only the first three pockets  from the table. We will let
$x$ be the number of $r\bar{r}rbg$ pockets, 
$y$ be the number of $b\bar{b}rbg$ pockets, and
$z$ be the number of $g\bar{g}rbg$ pockets so that,

\begin{equation}
         x+y+z=\eta.
\end{equation}
   
Since all particles in this system of $5\eta$ particles can interact with each other through color, there will be a total of $5\eta\left(5\eta-1\right)/2$ interactions. At the same time, we cannot be sure how the singlet is achieved in each pocket; we can only tell they are one of the three possibilities. This gives rise to $3^{\eta}$ possible configurations and, at any given time, the system would be found in one of those configurations. In order to give equal footing to all the color combinations, we counted all the possible interactions across all configurations. In this distribution, there will be $3^{\eta}5\eta\left(5\eta-1\right)/2$ occurances across the six possible color interactions. In order to simplify the calculation, we distribute the calculation into three different categories and put them together in the sub-subsections to follow.

\subsubsection{Occurrence of interactions between same type of pockets}
Here, we count the number of occurrences for interactions between the same type of pockets. For example, when a $r\bar{r}rbg$ pocket interacts with another $r\bar{r}rbg$ pocket, as given in Table~\ref{tab:tab3}, we can have six CCR type of interactions, ten CCA type of interactions, and so on, as shown in Table~\ref{tab:tab4}.
       
When $x$ number of $r\bar{r}rbg$ pockets interact with each other, the arrangement in Table~\ref{tab:tab4} gets repeated $x\left(x-1\right)/2$ times. For instance, the CCR type interaction occurs  $6x\left(x-1\right)/2$ times, the CCA type will occur $10 x\left(x-1\right)/2$ times, and so on. A similar argument can be made for a $b\bar{b}rbg$ pocket interacting with another $b\bar{b}rbg$ and a $g\bar{g}rbg$ pocket interacting with another $g\bar{g}rbg$, both of which will yield the same number as counted using Table~\ref{tab:tab3}. So, the total number of times the CCR type interactions can occur can be written as $6\left(x(x-1)+y(y-1)+z(z-1)\right)/2$, for the CCA type interactions we have $10 \left(x(x-1)+y(y-1)+z(z-1)\right)/2$, and so on. 

\begin{table}[ht]
\centering
\begin{tabular}[t]{lccccc}
\toprule
&$r$&$\bar{r}$&$r$&$b$&$g$\\
\midrule
$r$&$rr$&$r$$\bar{r}$&$rr$&$rb$&$rg$\\
$\bar{r}$&$\bar{r}r$&$\bar{r}\bar{r}$&$\bar{r}r$&$\bar{r}b$&$\bar{r}g$\\
$r$&$rr$&$r\bar{r}$&$rr$&$rb$&$rg$\\
$b$&$br$&$b\bar{r}$&$br$&$bb$&$bg$\\
$g$&$gr$&$g\bar{r}$&$gr$&$gb$&$gg$\\
\bottomrule\\
\end{tabular}
\caption{Possible quark interactions between the same type of pocket.}
\label{tab:tab3}
\end{table}%

\begin{table}[ht]
\centering
\begin{tabular}[t]{cc}
\toprule
Interaction &Number of Times\\
\midrule
CCR &6\\
CCA &10\\
CAR &4 \\
CAA &4 \\
AAR &1 \\
AAA &0 \\
\bottomrule\\
\end{tabular}
\caption{Occurrence of interactions between the same types of pockets}
\label{tab:tab4}
\end{table}%

\subsubsection{Occurrence of interactions between different pockets}
Here, we count the number of occurrences of interactions between different types of pockets. For example, when an $r\bar{r}rbg$ pocket interacts with a $b\bar{b}rbg$ pocket, as given in Table~\ref{tab:tab5}, we can have five CCR type interactions, eleven CCA type interactions, and so on. The complete list is shown in Table~\ref{tab:tab6}.

\begin{table}[ht]
\centering
\begin{tabular}[t]{lccccc}
\toprule
&$r$&$\bar{r}$&$r$&$b$&$g$\\
\midrule
$b$&$br$&$b$$\bar{r}$&$br$&$bb$&$bg$\\
$\bar{b}$&$\bar{b}r$&$\bar{b}\bar{r}$&$\bar{b}r$&$\bar{b}b$&$\bar{b}g$\\
$r$&$rr$&$r\bar{r}$&$rr$&$rb$&$rg$\\
$b$&$br$&$b\bar{r}$&$br$&$bb$&$bg$\\
$g$&$gr$&$g\bar{r}$&$gr$&$gb$&$gg$\\
\bottomrule\\
\end{tabular}
\caption{Possible quark interactions between two different types of pocket.}
\label{tab:tab5}
\end{table}%

\begin{table}[ht]
\centering
\begin{tabular}[t]{cc}
\toprule
Interaction &Number of Times\\
\midrule
CCR &5\\
CCA &11\\
CAR &6 \\
CAA &2 \\
AAR &0 \\
AAA &1 \\
\bottomrule\\
\end{tabular}
\caption{Occurrence of interactions for different types of pocket}
\label{tab:tab6}
\end{table}%

When $x$ number of $r\bar{r}rbg$ pockets interact with $y$ number of $b\bar{b}rbg$ pockets,  $y$ number of $b\bar{b}rbg$ pockets interact with $z$ number of $g\bar{g}rbg$ pockets, and $z$ number of $g\bar{g}rbg$ pockets interact with $x$ number of $r\bar{r}rbg$ pockets, CCR type of interaction occur  $5\left(xy+yz+zx\right)$ times, CCA type of interaction occur $11\left(xy+yz+zx\right)$ times and so on.

\subsubsection{Occurrence of interactions within each pocket}
The last case we need to consider are interactions within each pocket. We will first consider the pocket $r\bar{r} r b g$.
If we first consider $r$, it can interact with all of the other four quarks, namely, $r \bar{r}$, $rr$, $rb$ and $rg$. If we then consider the interactions involving $\bar{r}$, there are four possibilities, but only three unique interactions that we have not yet counted, namely, $\bar{r}r$, $\bar{r}b$ and $\bar{r}g$.  
Similarly, the two unique interactions involving the second $r$ that have not been previously counted are $rb$ and $rg$. Lastly, $b$ interacting with $g$ as $bg$ is the only remaining unique interaction and we have now counted all of the interactions. 
The results are shown in Table~\ref{tab:tab7} 
below.

\begin{table}[ht]
\centering
    \begin{tabular}[t]{cc}
        \toprule
        Interaction &Number of Times\\
        \midrule
            CCR &1\\
            CCA &5\\
            CAR &2 \\
            CAA &2 \\
            AAR &0 \\
            AAA &0 \\
        \bottomrule\\
    \end{tabular}
\caption{Interaction within the same pocket}
\label{tab:tab7}
\end{table}%

Exactly the same number of possibilities can be constructed for $b\bar{b}rbg$ as well as for $g\bar{g}rbg$ . Therefore, the total number of times CCR type of interaction can occur is $\eta$, CCA is $5\eta$ and so on. Here, we have used the fact that  $x+y+z=\eta$.

\subsubsection{Calculation of total number of occurrences}
In the next step we varied $x$ from 0 to $\eta$ and $y$ from 0 to $\left(\eta-x\right)$, thereby giving equal footing to all the color combinations and counted color
interactions. $E_i$ gives the total number of occurrences of the $i^{th}$ color interaction out of $3^\eta 5\eta(5\eta-1)/2$. Putting together what we obtained in previous sections, we have for $E_i$:  

    
    \begin{multline}E_i=\sum_{x=0}^\eta \sum_{y=0}^{\eta-x} \frac{\eta !}{x ! y ! (\eta-x-y) !} \Bigg[(x+y+z) \left( \begin{array}{c}
    1 \\
    5\\
    2\\
    2\\
    0\\
    0\\
    \end{array} \right) \\
    + \left(\frac{x(x-1)}{2}+\frac{y(y-1)}{2}+\frac{z(z-1)}{2}\right) \left( \begin{array}{c}
    6\\
    10\\
    4\\
    4\\
    1\\
    0\\
    \end{array} \right)
     +(xy+yz+zx) \left(\begin{array}{c}
    5\\
    11\\
    6\\
    2\\
    0\\
        1\\
    \end{array} \right) \Bigg]
    \label{Ei}
    .\end{multline}
Here, $E_i$ has been expressed in terms of vectors where the components of each of the column vectors denote the calculations summarized in Tables \ref{tab:tab4}, \ref{tab:tab6}, and \ref{tab:tab7}. The first term in $E_i$ corresponds to the interactions \textit{within a pocket}, the second term corresponds to interactions \textit{between the same type of pockets}, and the last corresponds to interactions \textit{between different pockets}. As an example, the first component of $E_i$ tells us the total number of occurrences of the $i^{th}$ color interaction involving CCR which can are made up by CCR interactions within a pocket, between the same type of pockets, and between different pockets.  

Employing {\it Mathematica}, Eq.~(\ref{Ei}) simplifies to

  \begin{equation}     
     E_i = \left[ \begin{array}{c}
    3^{\eta-1}\eta(8 \eta-5)\\
    3^{\eta-1}\eta(16 \eta-1)\\
    2 \times 3^{\eta-1}\eta(4 \eta-1)\\
    2 \times 3^{\eta-1}\eta(2 \eta+1)\\
    \frac{1}{2}3^{\eta-1}\eta(\eta-1)\\
    3^{\eta-1}\eta(\eta-1)\\
    \end{array} \right].
    \label{E2}
 \end{equation}
Since $E_i$ is the total number of occurrences of the $i^{th}$ color interaction, the probability of the $i^{th}$ interaction can be obtained by dividing Eq.~(\ref{E2}) by the total number of possible interactions between the color combinations we're interested in, $3^\eta 5\eta(5\eta-1)/2$.  
In addition, we define a probability notation that will help us in the calculation of energies. In this new table,  $i$ and $j$ refer to colors, $P$ refers to probability of interaction with no anti particle, $\overline{P}$ refers to interaction between one particle and one anti-particle, whereas $\overline{\overline{P}}$ refers to interaction between two anti-particles. We also divide by three to provide equal footing for each color. In this notation, we arrive at Table~\ref{tab:tabx}.
    
       \begin{table}[ht]
    \centering
    \begin{tabular}[t]{cc}
    \toprule
    Color probability symbol &Probability value\\
    \midrule
     $P_{ii}$ & $\displaystyle\frac{2(8\eta-5)}{45(5\eta-1)}$\\
     $P_{ij}$  & $ \displaystyle\frac{2(16\eta-1)}{45(5\eta-1)}$\\
    $\overline{\overline{P}}_{ii}$ & $\displaystyle\frac{4(4\eta-1)}{45(5\eta-1)}$ \\
    $\overline{\overline{P}}_{ij}$ & $\displaystyle\frac{4(2\eta+1)}{45(5\eta-1)}$ \\
    $\overline{\overline{P}}_{ii}$ & $\displaystyle\frac{(\eta-1)}{45(5\eta-1)}$ \\
    $\overline{\overline{P}}_{ij}$ &  $\displaystyle\frac{2(\eta-1)}{45(5\eta-1)} $\\
    \bottomrule\\
    \end{tabular}
    \caption{Probabilities due to color interactions in multi-pentaquarks.}
    \label{tab:tabx}
    \end{table}%
In Table~\ref{tab:tabx} it is understood that $i<j$ in $P_{ij},\overline{P}_{ij}$ and $\overline{\overline{P}}_{ij}$.

\subsection{Cross check on our counting}
As a cross check on the probabilities, adding them up yields,
    \[ \sum_{i \leqslant j}^3 P_{ij}+ \overline{P}_{ij}+ \overline{\overline{P}}_{ij}=1, \]
so the probabilities sum to one, as desired. 
    
We will also check our pentaquark model to ensure that it is indeed a color singlet. Let $\vec{Q}$ denote the total color charge of the quarks. By definition, we have,
    \[ \vec{Q}= \sum_{i=1}^{5\eta} \vec{q_i}. \]
    Squaring both the sides, 
    \begin{align}
        \vec{Q} \cdot \vec{Q} &= \sum_{i=1}^{5\eta}          q_i^2 + 2 \sum_{i \ne j} \vec{q_i} \cdot \vec{q_j} \\
        &= 5\eta \cdot \frac{4}{3}g^2 + 2 \times \frac{1}{3^\eta} \sum_i E_i C_i. 
    \end{align}
In the first term, $q_i \cdot q_j=\frac{4}{3}g^2$. In the second term, we have divided by $3^\eta$ to average over all the possible configurations. Here, $C_i$ is the coupling constant of $i^{th}$ interaction. Using {\it Mathematica}, we find that, 
\[ \sum_i E_i C_i = -10 \times 3^{\eta-1} g^2 \eta. \]
Therefore, using this value,
\[ \vec{Q} \cdot \vec{Q}= 0.\]
Hence, we see that our model is, indeed, a color singlet.

It should also be noted that, if we add the product of coupling and probability, we find that $-\frac{4}{3}g^2/(5\eta-1)$. Here, the negative sign indicates that the system is attractive because of the collective residual color coupling alone, even in absence of volume pressure.

\section{Flavors, Colors and Normalization Conditions}\label{flav}
In the previous section, we examined the system of $5\eta$ particles in terms of color. Now, we wish to examine the same system in terms of flavor and then combine our results with the corresponding color probabilities we calculated earlier. After that, we convert the discrete system into a continuum density of states and, finally, obtain normalization conditions that will help us calculate system energies.

\subsection{Counting based on flavors}
Our multi-quark system consists of $5\eta$ particles, where $\eta$ is the number of pockets. In each pocket there are four quarks and one antiquark. If $N_I$ and  $\bar{N}_I$ represent the number of flavors and the number of anti-flavors with degeneracy factors $g_I$ and $\bar{g_{I}}$, respectively, then,

    \begin{equation}
    \sum_I g_I N_I= 4\eta,
    \label{gini}
    \end{equation}
    \begin{equation}
    \sum_I \bar{g}_I \bar{N}_I=\eta,
    \label{gini2}
    \end{equation} 
where $I=1$ indicates light quarks and $I=2$ indicates heavy quarks.  Degeneracy factors $g$ and $\Bar{g}$ can take on a value of one, two, three or four depending on whether it is a light or heavy quark, which will be further explained in the results. 
    
\subsection{Putting flavors and colors together}\label{3.2}

Since there are $4\eta$ quarks and $\eta$ anti-quarks, we can expect $4\eta\left(4\eta-1\right)/2$ interactions between colors, $\eta\left(\eta-1\right)/2$ interactions between anti-colors, and $4\eta^2$ interactions between color and anti-color, all of which add up to $5\eta(5\eta-1)/2$. Interactions between two flavors can only be either CCA or CCR type, interactions between two anti-flavors can only be either AAR or AAA, and interactions between a flavor and anti-flavor can only be CAR or CAA. Using the equations above, we can summarize the interactions as those shown in Table~\ref{tab:tab8}. 

    \begin{table}[ht]
    \centering
    \begin{tabular}[t]{cc}
    \toprule
    Interactions &Number of Times\\
    \midrule
    CCR \& CCA & $ \displaystyle \sum_I {g_I N_I} \times \frac{1}{2}\left(\sum_J g_J N_J-1\right)$\\
    AAR \& AAA & $ \displaystyle \sum_I \bar{g}_I \bar{N}_I \times \frac{1}{2}\left(\sum_J \bar{g}_J \bar{N}_J-1\right)$\\
    CAR \& CAA & $\displaystyle \sum_{I,J} \bar{N}_I N_J \bar{g}_I g_J$ \\
    \bottomrule\\
    \end{tabular}
    \caption{Interactions due to flavor}
    \label{tab:tab8}
    \end{table}%

By assigning the flavors with color interaction probabilities, we can develop the expression for the potential energy. In the expression for potential energy in Eq.~(26) of Ref.~\cite{Baral:2019qpl}, there are terms related to CCR and CCA, terms for AAR and AAA, and one term for CAR and CAA type. They can simply be obtained by the expansion of the terms displayed above and separated into same flavor and different flavor pieces.
    
\subsection{Discrete to continuum system}
The system we have so far described has been one of discrete particles and, in order to apply the TF model, we have to convert it into a continuous system. If $n_i^I(r)$ and $\bar{n}_i^I(r)$ represent quark density of particles and anti-particles with flavor index $I$ and color index $i$, respectively, then Eq.~(\ref{gini}) and Eq.~(\ref{gini2}) can be written as

\begin{equation}
    \sum_{i,I} \int \text{d}^3r\, n_i^I(r) = {4\eta},
\end{equation}
and
\begin{equation}
    \sum_{i,I} \int \text{d}^3r\, \bar{n}_i^I(r)={\eta}.
\end{equation}
In the above equations, the degeneracy factors are already included in the quark densities $n_i^I(r)$ and $\bar{n}_i^I(r)$. For a particular color index $i$, the above equations can be written as,

\begin{equation}
    \sum_I \int \text{d}^3r\, n_i^I(r) = \frac{4\eta}{3},
\end{equation}
and
\begin{equation}
    \sum_I \int \text{d}^3r\, \bar{n}_i^I(r)=\frac{\eta}{3}.
\end{equation}
Similarly, for a particular flavor index $I$, these equations become,
\begin{equation}
    \sum_i \int \text{d}^3r\, n^I_i(r)= N_I g_I,
\label{439a}
\end{equation}
and 
\begin{equation}
    \sum_i \int \text{d}^3r\, \bar{n}_i^I(r)= \bar{N}_I \bar{g}_I .
\label{445a}
\end{equation}

\subsection{Fermi-Dirac normalization}
For individual colors, Eq.~(\ref{439a}) can be written as

\begin{equation}
        3 \int \text{d}^3r\, n^I_i(r) = N_I g_I.   \label{norm1}
    \end{equation}
We will assume equal quark color content and drop the index $i$ from this equation, remembering to sum over colors later. This gives
      
       \begin{equation}
        3 \int \text{d}^3r\, n^I (r) = N_I g_I .  
        \label{455a}
    \end{equation}
Similarly, for anti-particles, we have,
 \begin{equation}
    3 \int \text{d}^3r\, \bar{n}^I(r)= \bar{N}_I \bar{g}_I .
    \label{455b}
    \end{equation}
We will now introduce Thomas-Fermi functions, $f_I(r)$ and $\bar{f}_I(r)$, as
 \begin{equation}
        f_I(r)= \frac{ra }{2 \times \frac{4\alpha_s}{3}}\left(\frac{6 \pi^2 n^I(r)}{g_I} \right)^\frac{2}{3},
    \label{eq2}
    \end{equation}
and
    \begin{equation}
        \bar{f}_I(r)= \frac{ra}{2 \times \frac{4\alpha_s}{3}}\left(\frac{6 \pi^2 \bar{n}^I(r)}{\bar{g}_I} \right)^\frac{2}{3},
    \label{eq3}
    \end{equation}
where $a = \hbar/(m_1 c)$ gives the scale, $m_1$ is the mass of lightest quark, and $\alpha_s=g^2/(\hbar c)$ is the strong coupling constant. Note that $g_I$ and $\bar{g}_I$ are degeneracy factors. Eq.~(\ref{455a}) and Eq.~(\ref{455b}) can now be written as
    \begin{equation}
        \left( \frac{8 \alpha_s}{3a} \right)^\frac{3}{2} \frac{2}{\pi} \int_0^{r_{\text{max}}} \text{d}r \sqrt{r} \left( f_I(r) \right)^\frac{3}{2}=N_I, 
    \label{483d}
    \end{equation}
and
    \begin{equation}
        \left( \frac{8 \alpha_s}{3a} \right)^\frac{3}{2} \frac{2}{\pi} \int_0^{\bar{r}_{\text{max}}} \text{d}r \sqrt{r} \left( \bar{f}_I(r) \right)^\frac{3}{2}=\bar{N}_I. 
    \label{483b}
    \end{equation}
We will introduce a dimensionless parameter $x$ such that $r=Rx$ where, 
    
    \begin{equation}
    R=\left(\frac{a}{2 \times \frac{4 \alpha_s}{3}} \right) \left(\frac{3 \pi \eta}{2} \right)^\frac{2}{3}
    \label{rad}.
    \end{equation}
In terms of the dimensionless parameter $x$, Eq.~(\ref{483d}) and Eq.~(\ref{483b}) reduce to the following normalization conditions:

    \begin{equation}
        \int_0^{x_{\text{max}}} \text{d}x \sqrt{x} \left( f_I(x) \right)^\frac{3}{2} = \frac{N_I}{3\eta},
    \label{498a}
    \end{equation}
and

    \begin{equation}
    \int_0^{\bar{x}_{\text{max}}} \text{d}x \sqrt{x} \left( \bar{f}_I(x) \right)^\frac{3}{2} = \frac{\bar{N}_I}{3\eta}.
    \label{498s}
    \end{equation}

\section{Expression for Total Energy}\label{te}

In this section, we first introduce the TF model to show how the kinetic and potential energies are expressed as a function of the density of states. We then use the interaction probabilities from previous sections to build the expression for the potential energy for a family of multi-pentaquarks. The Thomas-Fermi Statistical model is a semi-classical model introduced to many fermion systems. It treats particles as a Fermi gas at $T = 0$. Despite it utilizing Fermi statistics, it is not fully quantum mechanical since it does not have a quantum mechanical wave function but, rather, a central function related to particle density. This function is determined by filling states up to the Fermi surface at each physical location. The key idea of the Thomas-Fermi quark model is to express both the kinetic energy and the attractive and repulsive potential energy contributions as a simple function of quark density. 
    
The general expression for the kinetic energy is explained in Section 2 of Ref.~\cite{Baral:2019qpl} and is embodied in Eq.~(25)\footnote{Please note the incorrect numerator in Eq.~(25) of Ref.~\cite{Baral:2019qpl}. The correct numerators should replace the 6$\pi^2$ factors with 2$\pi^2$. This error propagates to Eq.~(27), but no further. In addition, the kinetic energy terms in Eqs.~(28) and (31) should not have the number densities $N_I$ and $\bar{N}_I$ in their numerators. This error also does not propagate.} of that reference. In order to apply this expression to the quark system, we will introduce normalized degeneracy densities $ \hat{n}_i^I$ and $\hat{\bar{n}}_i^I$ given by
    \begin{equation}
        \hat{n}_i^I= \frac{3 n^I_i}{N_I} ,
    \end{equation}
and
    \begin{equation}
     \hat{\overline{n}}_i^I= \frac{3 \overline{n}_i^I}{\overline{N}_I}. 
     \end{equation}
This new form of the quark density is helpful in correctly normalizing energies when continuum sources are used. When summed over flavors and colors, this yields the total kinetic energy ($T$)
     \begin{equation}
        T=\sum_{i,I} \int^{r_{\text{max}}} \text{d}^3r \frac{\left(2 \pi^2 \hbar^3 N_I \hat{n}_i^I(r)\right)^\frac{5}{3}}{20 \pi^2 \hbar^3 m_I (g_I)^\frac{2}{3}} + \sum_{i,I} \int^{\bar{r}_{\text{max}}} \text{d}^3r \frac{\left(2 \pi^2 \hbar^3 \bar{N}_I \hat{\bar{n}}_i^I(r)\right)^\frac{5}{3}}{20 \pi^2 \hbar^3 \bar{m}_I (\bar{g}_I)^\frac{2}{3}}.
        \label{kin} 
    \end{equation}
   
The procedure for determining the potential energy ($U$) will mirror that of \cite{Baral:2019qpl} in Eq.~(26) with the exception of the new probabilities found for the multi-pentaquarks. We use the probabilities and flavor statistics of the previous section, giving
   
\begin{multline*}U= \frac{4}{3}g^2 \sum_I \left(  \frac{N_I(N_I-1)}{2} + \frac{N_I g_I(g_I-1)}{2(g_I)^2}   \right) \\
\times \int \int \text{d}^3r \,\text{d}^3r' \frac{\left(\sum_i P_{ii} \hat{n}_i^I(r)\hat{n}_i^I(r')-\frac{1}{2} \sum_{i<j} P_{ij} \hat{n}_i^I(r) \hat{n}_j^I(r') \right)}{|\vec{r}-\vec{r}\,'|}\\
+ \frac{4}{3}g^2 \sum_{I \ne J} \frac{N_I N_J}{2} \int \text{d}^3r \,\text{d}^3r' \int \frac{\left( \sum_i P_{ii} \hat{n}_i^I(r)\hat{n}_i^J(r')-\frac{1}{2} \sum_{i<j} P_{ij} \hat{n}_i^I(r) \hat{n}_j^J(r') \right)}{|\vec{r}-\vec{r}\,'|} \end{multline*}
\begin{multline}+ \frac{4}{3}g^2 \sum_{I} \left(\frac{\overline{N}_I (\overline{N}_I-1)}{2} + \frac{\overline{N}_I \overline{g}_I(\overline{g}_I-1)}{2(\overline{g}_I)^2}  \right)\\
\times \int \int \text{d}^3r \,\text{d}^3r' \frac{\left( \sum_i \overline{\overline{P}}_{ii} \hat{\overline{n}}_i^I(r)\hat{\overline{n}}_i^I(r')-\frac{1}{2} \sum_{i<j} \overline{\overline{P}}_{ij} \hat{\overline{n}}_i^I(r) \hat{\overline{n}}_j^I(r') \right)}{|\vec{r}-\vec{r}\,'|} \\
+ \frac{4}{3}g^2 \sum_{I \ne J} \frac{\overline{N}_I \overline{N}_J}{2} \int \text{d}^3r\, \text{d}^3r'\int \frac{ \left( \sum_i \overline{\overline{P}}_{ii} \hat{\overline{n}}_i^I(r)\hat{\overline{n}}_i^J(r')-\frac{1}{2} \sum_{i<j} \overline{\overline{P}}_{ij} \hat{\overline{n}}_i^I(r) \hat{\overline{n}}_j^J(r') \right)}{|\vec{r}-\vec{r}\,'|} \\
  - \frac{4}{3}g^2 \sum_{I, J} \overline{N}_I N_J \int \int \text{d}^3r \,\text{d}^3r' \frac{\left( \sum_i \overline{P}_{ii} \hat{\overline{n}}_i^I(r)\hat{n}_i^J(r')-\frac{1}{2} \sum_{i<j} \overline{P}_{ij} \hat{\overline{n}}_i^I(r) \hat{n}_j^J(r') \right)}{|\vec{r}-\vec{r}\,'|}.
		   \label{pot}
 \end{multline} 
We have	
\begin{equation}
   P_{ii}-\frac{1}{2}P_{ij}= -\frac{3}{5(5\eta-1)},
    \end{equation}
    \begin{equation}
    \overline{\overline{P}}_{ii}-\frac{1}{2}\overline{\overline{P}}_{ij}=0,
    \end{equation}
    and
    \begin{equation}
        \overline{P}_{ii}-\frac{1}{2}\overline{P}_{ij}=\frac{2}{5(5\eta-1)}.
        \end{equation} 
Furthermore, with equal weighting provided to all the colors, we arrive at the final expression for the total energy ($E$),  
                	
\begin{multline} E= \sum_{I} \int^{r_{\text{max}}} \text{d}^3r \frac{3 \left( 6 \pi^2 \hbar^3 \right)^\frac{5}{3}}{20 \pi^2 \hbar^3 m_I (g_I)^\frac{2}{3}} \left(n^I(r)\right)^\frac{5}{3}  + \sum_{I} \int^{r_{\text{max}}} \text{d}^3r \frac{3 \left( 6 \pi^2 \hbar^3 \right)^\frac{5}{3}}{20 \pi^2 \hbar^3 \bar{m}_I (\bar{g}_I)^\frac{2}{3}} \left(\bar{n}^I(r)\right)^\frac{5}{3} \\
 - \frac{18 g^2}{5(5\eta-1)}\sum_I \frac{(g_I N_I-1)}{g_I N_I}\int \int \text{d}^3r \,\text{d}^3r' \frac{n^I(r) n^I(r')}{|\vec{r}-\vec{r}\,'|}\\
 - \frac{18 g^2}{5(5\eta-1)}\sum_{I \ne J}\int \int \text{d}^3r \,\text{d}^3r' \frac{n^I(r) n^J(r')}{|\vec{r}-\vec{r}\,'|} \\
 - \frac{24 g^2}{5(5\eta-1)}\sum_{I, J}\int \int \text{d}^3r \,\text{d}^3r'\frac{\bar{n}^I(r) n^J(r')}{|\vec{r}-\vec{r}\,'|} ,\\
 \end{multline}
 where we have switched to the single-color particle densities $n^I$ and $\bar{n}^I$ with normalizations (\ref{455a}) and (\ref{455b}).

\section{Thomas-Fermi Pentaquark Equations}\label{tfe}

In Section \ref{te}, we calculated the total energy of a family of pentaquarks. Now, we want to formally minimize the energy by varying the particle densities, while keeping the quark number constant. This will give us the differential equations we need to solve. 

Let's introduce Lagrange undetermined multipliers $\lambda^I$ and $\bar{\lambda}^I$ associated with the constraints,

    \begin{equation}
        3\int \text{d}^3r\, n^I(r)=N_Ig_I,
    \end{equation}
and
    \begin{equation}
        3\int \text{d}^3r\,\bar{n}_I(r)=\bar{N}_I \bar{g}_I,
    \end{equation}
respectively. Then, the total energy becomes

\begin{multline}E= \sum_{I} \int^{r_{\text{max}}} \text{d}^3r \frac{3 \left( 6 \pi^2 \hbar^3 \right)^\frac{5}{3}}{20 \pi^2 \hbar^3 m_I (g_I)^\frac{2}{3}} \left(n^I(r)\right)^\frac{5}{3} 
+ \sum_{I} \int^{r_{\text{max}}} \text{d}^3r \frac{3 \left( 6 \pi^2 \hbar^3 \right)^\frac{5}{3}}{20 \pi^2 \hbar^3 \bar{m}_I (\bar{g}^I)^\frac{2}{3}} \left(\bar{n}^I(r)\right)^\frac{5}{3}\\
- \frac{18 g^2}{5(5\eta-1)}\sum_I \frac{(g_I N_I-1)}{g_I N_I}\int \int \text{d}^3r \,\text{d}^3r' \frac{n^I(r) n^I(r')}{|\vec{r}-\vec{r}\,'|} \\
- \frac{18 g^2}{5(5\eta-1)}\sum_{I \ne J}\int \int \text{d}^3r \,\text{d}^3r'\frac{n^I(r) n^J(r')}{|\vec{r}-\vec{r}\,'|}
- \frac{24 g^2}{5(5\eta-1)}\sum_{I, J}\int \int \text{d}^3r \,\text{d}^3r' \frac{\bar{n}^I(r) n^J(r')}{|\vec{r}-\vec{r}\,'|} \\
+ \sum_I \lambda^I \left( 3 \int^{r_{\text{max}}} \text{d}^3r\, n^I(r)-g_I N_I \right) + \sum_I \bar{\lambda}_I \left(3 \int^{r_{\text{max}}} \text{d}^3r\, \bar{n}^I(r)- \bar{N}_I \bar{g}_I \right).
\label{eqn}	 
 \end{multline}
Once again, the purpose of adding these terms involving the Lagrange multipliers is to allow a minimization of the total energy while keeping particle number fixed. 
    
The variation of the density $\delta n^I(r)$ in Eq.~(\ref{eqn}) gives
    \begin{multline} \frac{\left(6 \pi^2 \hbar^3\right)^\frac{5}{3}}{\pi^2 \hbar^3} \frac{1}{4 m_I} \left(\frac{n^I(r)}{g_I}\right)^\frac{2}{3} =\frac{18 g^2}{5(5\eta-1)} \sum_{I} \frac{N_I g_I-1}{N_I g_I}\int^{r_{\text{max}}} \text{d}^3r'\frac{ n^I(r')}{|\vec{r}-\vec{r}\,'|} \\
        + \frac{18 g^2}{5(5\eta-1)} \sum_{I \ne J}\int^{r_{\text{max}}} \text{d}^3r'\frac{n^J(r')}{|\vec{r}-\vec{r}\,'|}  + \frac{24g^2}{5(5\eta-1)} \sum_{I}\int^{r_{\text{max}}} \text{d}^3r'\frac{\bar{n}^J(r')}{|\vec{r}-\vec{r}\,'|} -3 \lambda^I. 
    \label{eq1}
    \end{multline}
Similarly, variation of the density $\delta \bar{n}^I(r)$ in Eq.~(\ref{eqn}) gives,
    \begin{equation}
        \frac{\left(6 \pi^2 \hbar^3\right)^\frac{5}{3}}{\pi^2 \hbar^3} \frac{1}{4 \bar{m}_I} \left(\frac{\bar{n}^I(r)}{\bar{g}_I}\right)^\frac{2}{3}=\frac{24g^2}{5(5\eta-1)} \sum_{I}\int^{r_{\text{max}}} \text{d}^3r'\frac{ n^J(r')}{|\vec{r}-\vec{r}\,'|} -3 \bar{\lambda}^I.
        \label{eq5} 
    \end{equation}
We also know that, 
    \begin{equation}
        \int^{r_{\text{max}}} \text{d}^3r'\frac{n^J(r')}{|\vec{r}-\vec{r}\,'|} =4 \pi \left[ \int^r_0 \text{d}r' r'^2 \frac{n^J(r')}{r} + \int^{r_{\text{max}}} \text{d}r' r'^2 \frac{n^J(r')}{r'} \right].
    \label{eq4}
    \end{equation}
Let us define $\bar{\alpha}_I \equiv \bar{m}_1/\bar{m}_I$ as the ratio of mass of the lightest quark to the $I^{th}$ flavor quark. Combining Eqs.~(\ref{eq2}), (\ref{eq3}), and (\ref{eq4}) with Eq.~(\ref{eq1}), we obtain in terms of the dimensionless parameter $x$,
    \begin{multline}   
         \alpha_I f_I(x)= -\frac{\lambda^I}{\frac{4}{3}g^2}Rx
         + \frac{6\eta}{5(5\eta-1)} \sum_I \bar{g}_I \left[ \int_0^x \text{d}x' \sqrt{x'} \bar{f}_I(x')+ x\int_x^{x_{\text{max}}} \text{d}x' \frac{\left(\bar{f}_I(x')\right)^\frac{3}{2}}{\sqrt{x'}}\right]  
    \\
    + \frac{9\eta}{10(5\eta-1)} \Bigg\{ \frac{(N_Ig_I-1)}{N_I} \left[ \int_0^x \text{d}x' \sqrt{x'}f_I(x') + x\int_x^{x_{\text{max}}} \text{d}x' \frac{\left(\bar{f}_I(x')\right)^\frac{3}{2}}{\sqrt{x'}} \right]
    \\
        + \sum_{I \ne J} g_J \left[ \int_0^x \text{d}x' \sqrt{x'}f_J(x') + x\int_x^{x_{\text{max}}} \text{d}x' \frac{\left(\bar{f}_J(x')\right)^\frac{3}{2}}{\sqrt{x'}} \right] \Bigg\}. 
        \label{deri}
    \end{multline}
Differentiating Eq.~(\ref{deri}) twice, we get the first of two Thomas-Fermi differential equations, namely,
    \begin{equation}
        \alpha_I \frac{\text{d}^2 f_I(x)}{\text{d}x^2}=-\frac{6\eta}{5(5\eta-1)}\sum_I \bar{g}_I \frac{\left( \bar{f}_I(x) \right)^\frac{3}{2}}{\sqrt{x}} - 
        \frac{9\eta}{10(5\eta-1)}\left[\frac{(N_I g_I-1)}{N_I} \frac{\left( f_I(x) \right)^\frac{3}{2}}{\sqrt{x}} +\sum_{I \ne J} g_J \frac{\left( f_J(x) \right)^\frac{3}{2}}{\sqrt{x}} \right].
    \label{700a}
    \end{equation}
Similarly, combining Equations (\ref{eq2}), (\ref{eq3}), and (\ref{eq4}) with (\ref{eq5}) and using the dimensionless parameter $x$, we have
\begin{equation}
\bar{\alpha}_I \bar{f}_I(x)=-\frac{\bar{\lambda}^I}{\frac{4}{3}g^2}Rx 
        + \frac{6 \eta}{5(5\eta-1)} \sum_J g_J \left[ \int_0^x \text{d}x' \sqrt{x'} \left(f_J(x') \right)^\frac{3}{2} + x \int^{x_{\text{max}}}_x \text{d}x' \frac{\left(f_J(x')\right)^\frac{3}{2}}{\sqrt{x'}} \right].
    \label{eq}
 \end{equation}
Finally, differentiating Eq.~(\ref{eq}) twice yields, 
    \begin{equation}
    \bar{\alpha}_I \frac{\text{d}^2\bar{f}_I(x)}{\text{d}x^2}= -\frac{6\eta}{5(5\eta-1)} \sum_J g_J \frac{(f_J(x))^\frac{3}{2}}{\sqrt{x}}.
    \label{700b}
    \end{equation}
Eq.~(\ref{700a}) and Eq.~(\ref{700b}) are Thomas-Fermi differential equations for our system of pentaquarks.

\section{Consistency Conditions and Parameters}\label{ccp}
Similar to the atomic model, the TF quark model assumes heavy particles in the central region and light particles spread outside of it. In the case of hidden charm multi-pentaquarks, the $u$ and $d$ quarks are light particles which can have larger radii while the $c$ and $\bar{c}$ are the heavy quarks relative to the $u$ and $d$ quarks. In this paper, we will be investigating whether the $c$ or the $\bar{c}$ will be found within the innermost radius. This can depend on several factors like color-coupling probabilities, strength of color-coupling, separation between particles, number of flavors, mass of flavors, and more. 

In this section, we first obtain the consistency conditions required for our model to give a single collective density of states and then discuss how the parameters will be chosen to solve consistency conditions numerically.

\subsection{The consistency conditions}
For the heavy particles, the TF function is $f_2(x)$ and for the heavy antiparticles $\bar{f}_2(x)$. For the light quarks, the TF function is $f_1(x)$ and for light anti-particles, $\bar{f}_1(x)$. Since we are not dealing with light anti-particles, $\bar{f}_1(x)$ will be zero.

The TF quark model creates a single, collective, spherically-symmetric density of states. So, we will assume that the TF fermi functions of all particles will be linearly related to each other by some proportionality factor $k$ and $\bar{k}$. In other words, 
\begin{equation}
f_1(x)=kf_2(x)
\label{kfirst},
\end{equation}
\begin{equation}
\bar{f}_2(x)=\bar{k}f_2(x).
\label{kseco}
\end{equation} 
Using these TF functions in the TF differential equations, we obtain,

\begin{equation}
\bar{\alpha}_2 \frac{\text{d}^2\bar{f}_2}{\text{d}x^2}= -\frac{6\eta}{5(5\eta-1)\sqrt{x}}\left[g_1 \left(f_1(x)\right)^\frac{3}{2}+g_2\left(f_2(x)\right)^\frac{3}{2} \right]
\label{second},
\end{equation}
\begin{equation}
\alpha_1 \frac{\text{d}^2f_1}{\text{d}x^2}= -\frac{6\eta}{5(5\eta-1)\sqrt{x}}\bar{g}_2 \left(\bar{f}_2\right)^\frac{3}{2} \\
-\frac{9\eta}{10(5\eta-1)\sqrt{x}} \left[\frac{(N_1 g_1-1)}{N_1} \left(f_1(x)\right)^\frac{3}{2}+g_2\left(f_2(x)\right)^\frac{3}{2} \right],
\label{third} 
\end{equation}
and
\begin{equation}
\alpha_2 \frac{\text{d}^2f_2}{\text{d}x^2}= -\frac{6\eta}{5(5\eta-1)\sqrt{x}}\bar{g}_2 \left(\bar{f}_2\right)^\frac{3}{2} \\
-\frac{9\eta}{10(5\eta-1)\sqrt{x}} \left[\frac{(N_2 g_2-1)}{N_2} \left(f_2(x)\right)^\frac{3}{2}+g_1\left(f_1(x)\right)^\frac{3}{2} \right]
\label{fourth}.
\end{equation}
Inserting Eqs.~(\ref{kfirst}) and (\ref{kseco}) into Eq.~(\ref{second}), (\ref{third}), and (\ref{fourth}), we obtain,
\begin{equation}
\frac{\text{d}^2 f_2(x)}{\text{d}x^2}= Q \frac{\left(f_2(x)\right)^\frac{3}{2}}{\sqrt{x}}
\label{x2},
\end{equation}
where,
\begin{equation}
Q= -\frac{6\eta}{5(5\eta-1) \bar{\alpha}_2 \bar{k}} \left( g_1 k^\frac{3}{2} + g_2 \right)
\label{con1},
\end{equation}
\begin{equation}
Q= -\frac{6\eta}{5(5\eta-1)k} \left(\bar{g}_2 \bar{k}^\frac{3}{2} + \frac{3}{4}\frac{(N_1 g_1-1)}{N_1} k^\frac{3}{2} + \frac{3}{4} g_2 \right),
\label{con2}
\end{equation}and
\begin{equation}
	Q= -\frac{6\eta}{5(5\eta-1)\alpha_2} \left(\bar{g}_2 \bar{k}^\frac{3}{2}+ \frac{3}{4}\frac{(N_2 g_2-1)}{N_2} + \frac{3}{4} g_1 k^\frac{3}{2} \right)
\label{con3}.
\end{equation}
Eqs.~(\ref{con1}), (\ref{con2}), and (\ref{con3}) are the consistency conditions. Here, $k$, $\bar{k}$ and $Q$ are three unknowns which can be calculated using these three consistency conditions.

\subsection{Methods and parameters}

The internal number parameters which enter the model are the particle state numbers $N_1, N_2$ and the antiparticle state number $\bar{N}_2$. In addition, there are the particle and antiparticle degeneracy factors $g_1, g_2$ and $\bar{g}_2$. These parameters satisfy the particle number constraints (\ref{gini}) and (\ref{gini2}). Due to their masses being nearly equal, we will assume the density functions of the $u$ and $d$ quarks to be the same. Furthermore, since spin up and spin down states are distinguishable between flavors, this gives us four distinguishable particles with the same mass. The degeneracy factor, $g_1$, for the light particle therefore will have a value from one to four. Charm and anti-charm can have a maximum degeneracy of two because there is no other quark with a similar mass.

In the following we will investigate the solution and interpretation of the consistency conditions in the special case of the ground state system for a given multi-pentaquark. We will therefore choose the maximum possible values for degeneracy factors $g_1, g_2$ and $\bar{g}_2$ for each value of $\eta$. We will also specialize to the equal heavy state and number situation: $N_2=\bar{N}_2$ and $g_2=\bar{g}_2$. Let us analyze the situation.

For $\eta=1$, the quark combination is $c\bar{c}qqq$, where $q$ represents a light quark species. If one $q$ is flavor $u$ with spin up, another $q$ is flavor $u$ with spin down, and the last $q$ is flavor $d$ with either spin up or in spin down state, all three light flavors are distinguishable. Therefore, there should be a degeneracy of three for the light quarks. Similarly, $c$ and $\bar{c}$ could either be spin up or spin down, so their degeneracy is one. Hence, we should have  
$N_1 = 1$,
$g_1 = 3$,
$N_2 = 1$,
$g_2 = 1$,
 $\bar{N}_2= 1$
and, $\bar{g}_2= 1$.
For $\eta=2$,
the quark combination is ($c\bar{c}qqq$ $c\bar{c}qqq$). In order to maximize the degeneracy factors, it is clear that the ground state is represented by 
$N_1 = 2$,
$g_1 = 3$,
$N_2 = 1$,
$g_2 = 2$,
 $\bar{N}_2= 1$ and,
$\bar{g}_2= 2$.
For $\eta=3$, the only state parameters which add up correctly are: 
$N_1 = 3$,
$g_1 = 3$,
$N_2 = 3$,
$g_2 = 1$,
 $\bar{N}_2= 3$,
$\bar{g}_2= 1$.
For $\eta=4$, the maximum degeneracy state should have $g_2=\bar{g}_2=2$. Then, maximizing the light quark degeneracy factor, we then expect the ground state to be given by $N_1 = 3$ and $g_1=4$. For $\eta=5$ the only possibility is
$N_1 = 5$,
$g_1 = 3$,
$N_2 = 5$,
$g_2 = 1$,
 $\bar{N}_2= 5$, and
$\bar{g}_2= 1$.
Finally, for $\eta=6$, we have the maximally degenerate state
$N_1 = 6$,
$g_1 = 3$,
$N_2 = 3$,
$g_2 = 2$,
 $\bar{N}_2= 3$ and
$\bar{g}_2= 2$.

\section{Results and Discussion}\label{rd}
The normalization condition from Eq.~(\ref{498a}) and Eq.~(\ref{498s}) can be written  for heavy charm and anti-charm as,
 \begin{equation}
 \int_0^{x_{\text{max}}} \text{d}x \sqrt{x} \left( f_2(x) \right)^\frac{3}{2} = \frac{N_2}{3\eta},
 \label{res1}
\end{equation}
and
  \begin{equation}
 \int_0^{\bar{x}_{\text{max}}} \text{d}x \sqrt{x} \left( \bar{f}_2(x) \right)^\frac{3}{2} = \frac{\bar{N}_2}{3\eta}
 \label{res2}.
\end{equation}
In the family of multi-pentaquarks under consideration, charm and anti-charm are always equal in number, hence $N_2=\bar{N}_2$. If we assume that $\bar{x}_{\text{max}}<x_{\text{max}}$ we may substitute
$\bar{f}_2(x)=\bar{k}f_2(x)$ for all $x$ in Eq.~(\ref{res2}). Thus, we find
\begin{equation}
 (\bar{k})^{3/2}\int_0^{\bar{x}_{\text{max}}} \text{d}x \sqrt{x} \left( f_2(x) \right)^\frac{3}{2} = \frac{N_2}{3\eta}.
 \label{res22}
\end{equation}
Since Eq.~(\ref{res1}) and Eq.~(\ref{res22}) have same right hand sides, it follows that the left hand sides should also be equal. Note that the function $\left( f_2(x) \right)^\frac{3}{2}$ is non-negative as it represents the number density and further that $x$,  $\bar{x}_{\text{max}}$ and $x_{\text{max}}$ are positive numbers. This implies that the value of integration keeps decreasing as the upper limit of integration decreases. In other words, if $\bar{x}_{\text{max}}<x_{\text{max}}$ then consequently $\bar{k}>1$ and vice-versa as this argument can be repeated by substitution in (\ref{res1}) rather than (\ref{res2}).

We solved the consistency conditions given by Eqs.(\ref{con1}), (\ref{con2}), and (\ref{con3}) using {\it Mathematica}. We used the mass of charm and anti-charm from \cite{Baral:2019qpl}; the masses of charm and anti-charm were 1553 MeV while the mass of the light quark was 306 MeV. We then obtained real values for $k$ and $\bar{k}$, the results of which are tabulated in Table~\ref{tab:tab9} below. 

\begin{table}[ht]
\centering
\begin{tabular}[t]{ccccccccc}
\toprule
$ \eta$& $N_1$ & $g_1$& $N_2$& $g_2$& $\bar{N}_2$& $\bar{g}_2$& $k$ &$\bar{k}$\\
\midrule
1 & 1 & 3 & 1 & 1 & 1 & 1 &0.286177&1.03934 \\
2 & 2 & 3 & 1 & 2 & 1 & 2 &0.24862  &0.882622\\
3 & 3 & 3 & 3 & 1 & 3 & 1 &0.225637&0.859367\\
4 & 3 & 4 & 2 & 2 & 2 & 2 &0.220509&0.823719\\
5 & 5 & 3 & 5 & 1 & 5 & 1 &0.213978&0.825001 \\
6 & 6 & 3 & 3 & 2 & 3 & 2 &0.213252&0.794695 \\
\bottomrule\\
\end{tabular}
\caption{Value of $k$ and $\bar{k}$ for various TF pentaquark states.}
\label{tab:tab9}
\end{table}%
\begin{figure}
\centering
\includegraphics[trim={1cm -1cm 1cm 1cm},clip,width=0.40\textwidth]{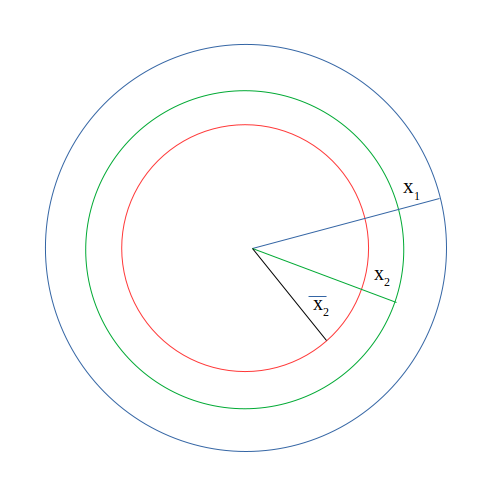}
\caption{For $\eta=1$, all three particles can be found inside the inner region represented by coordinate $\overline{x}_2$. The middle region bounded by coordinate $\overline{x}_2$ and $x_2$ is for heavy charm and light quarks but not anti-charm. The outer region between ${x}_2$ and  $x_1$ is populated only by light quarks.}
\label{fig:region_new}
\end{figure}
\begin{figure}
\centering
\includegraphics[trim={1cm 0cm 1cm 1.5cm},clip,width=0.46\textwidth]{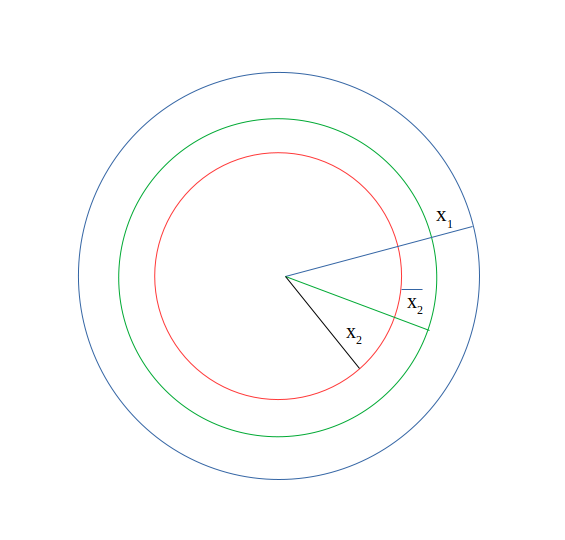}
\caption{For $\eta=\left(2, 3, 4, 5, 6\right)$ the region inside coordinate ${x}_2$ is populated by all three particles. The middle region bounded by coordinate  ${x}_2$ and $\overline{x}_2$ is for heavy anti-charm and light quarks but not charm. The outer region between $\overline{x}_2$ and  $x_1$ is populated only by light quarks.}
\label{fig:region_new1}
\end{figure}

We can see that $\bar{k}>1$ only for $\eta=1$. This implies that the anti-charm quark has the smallest radius for $\eta=1$ while the charm has the smallest radius for $\eta>1$. Furthermore, the value of $\bar{k}$ gets smaller with higher quark content. This suggests that the anti-charm spreads out further as the quark content of the system increases. However, a numerical statement will have to wait until an actual evaluation of the $f_2(x)$ function can be made. Note also the slight nonmonotonic behavior in $\bar{k}$ for $\eta = 4, 5$ and 6. The reason for this is believed to be due to $\eta = 4$ being the only configuration listed with $g_1=4$, and thus somewhat exceptional. It is also physically clear that the light quark TF function surrounds the charm and anti-charm regions.

In our further investigations of this family of pentaquarks, we will need three TF functions for light quarks, heavy charms and heavy anti-charms. The position of $x_{2}$ and $\bar{x}_2$ have to be chosen carefully, depending on the number of quarks. Figs.~\ref{fig:region_new} and \ref{fig:region_new1} illustrate the idea. 

\section{Future Work}\label{future}

There are additional steps which need to be taken before the mass spectrum and radial structure of multi-pentaquark family states can be fully delineated. The complete kinetic and potential energies will need to be developed and related to the TF functions $f_1, f_2$ and $\bar{f}_2$, with the understanding that $\bar{x}_2\ne x_2$ in general. The volume energy needs to be added and the numerical job of minimizing the total energy while solving the TF equations still remains to be done. 

The beginning point for the phenomenology and fits in this model is the LHCb states identified as pentaquarks with minimal quark content $c\bar{c} uud$; see the summary in Ref.~\cite{Zyla:2020zbs}. Note that a very simple model for hidden charm pentaquarks has a spin interaction Hamiltonian given by
\begin{equation}
\begin{aligned}
H_{\text{spin}}= \kappa_1 (s_1(s_1+1)-9/4) &+ \kappa_2 (s_2(s_2+1)-3/2) + \kappa_3 (J(J+1)\\&
-s_1(s_1+1) -s_2(s_2+1)),\label{spineqn}
\end{aligned}
\end{equation}
where $s_1$ is the light quark spin, $s_2$ is the heavy quark spin and $J$ is the total spin. The three terms represent the light-light, heavy-heavy and light-heavy spin interactions, respectively. We would expect that $\kappa_2<\kappa_1, \kappa_3$ based on the quark masses. It will be necessary to form a hypothesis on the spin and parity content of the LHCb pentaquark states in order to fit the model parameters, which include the charm and light quark masses, the strong interaction constant $\alpha_s$ and the bag parameter, $B$. On the other hand, this process has already been completed in the considerations of Ref.~\cite{Baral:2019qpl}, and we could proceed with the previous parameter set. Ideally, the use of two sets of parameters could give an indication of the systematic error in the model. In addition, to bring the evaluations to the same level of completeness as in Ref.~\cite{Liu:2012my,liu2019thomas}, the spin energies also need to be calculated for the various degenerate states based upon their nonrelativistic wavefunctions.

There have been model attempts to interpret pentaquarks as either a system made of diquarks\cite{Maiani:2015vwa} or triquarks\cite{Lebed:2015tna,Zhu:2015bba}, a loosely bound molecular model consisting of a charmed baryon and an anti-charmed meson\cite{Karliner:2015ina}, or compact hadro-charmonium states\cite{Sibirtsev:2005ex,Dubynskiy:2008mq,Li:2013ssa,Eides:2019tgv}. Our statistical model does not assume smaller subsytems of quarks {\it per se}, but postulates that the interactions between the quarks can be characterized by averaged gluonic color interactions. In the emergent system formed the charmed and anti-charmed quarks reside at the center of the system, and in this sense more closely resembles the structure associated with the hadro-charmonium picture. However, note that Eq.~(\ref{spineqn}) will produce a seven-plet of spin energy levels, the same as the molecular picture\cite{PhysRevLett.122.242001}. Ultimately, it will be the the comparison with the pentaquark energy levels, which are near $\Sigma_c^+ \bar{D}^0, \Sigma_c^+ \bar{D}^{0*}$ particle thresholds, which will be the most revealing. We now have all the tools to make this comparison: the baryon results from Refs.~\cite{Liu:2012my,liu2019thomas} (which need to be extended to charm quarks) and the meson results from Ref.~\cite{Baral:2019qpl} will allow us to calculate both thresholds and pentaquark energy levels when a unified set of parameters are used. We then use the reliable predictions from Thomas-Fermi particle density theory as a basis to search for families of such states. The ground states will only be stable if the average energy per quark is a decreasing function of quark number. We have found such a decrease for a type of meson state that we called Case 2 charmed and bottom systems in Ref.~\cite{Baral:2019qpl}. This is what we will search for in pentaquark families as well.

Although this represents a considerable amount of additional work, we are encouraged by the consistent mathematical structure and physical picture that seems to be emerging, and we pause here before continuing on.

\section{Conclusions and Acknowledgements}\label{conclusions}

We have initiated research into hidden charm multi-pentaquark families using the TF statistical model. We were able to evaluate the various color interaction terms and obtain the probabilities of particle and antiparticle interactions. This allowed us to form the kinetic and potential energies subject to flavor normalization conditions. We then obtained the appropriate TF differential equations and solved the consistency conditions for a number of multi-pentaquark ground states. We found that this required at least three TF functions with three different radii. We observed that for a pentaquark, the ordering of the quark radii is as in Fig.~\ref{fig:region_new} where the heavy charm antiquark is limited to the innermost region. As the quark content increases, however, the heavy charm, rather than the anti-charm, is limited to the center region, as in Fig.~\ref{fig:region_new1}. Our evaluations were carried out to $\eta=6$, i.e., for a state that is a combination of 6 pentaquarks.

We thank the Baylor University Research Committee, the Baylor Graduate School, and the Texas Advanced SuperComputing Center for partial support. We would like to acknowledge Mr.~Sujan Baral and Ms.~Pratigya Gyawali, CEO and COO respectively from Everest Institute of Science and Technology for initiating EVIST research collaborations. We thank Mr.~Bikram Pandey, Mr.~Shankar Parajuli and Mr.~Pravesh Koirala for partial calculations and other helpful considerations. We also acknowledge the Grant Office at Niagara County Community College as well as the National Science and Research Society for Educational Outreach of Nepal.

\vfill

\eject

\bibliography{referfile3}

\vfill

\eject

\end{document}